\documentclass{llncs}
\usepackage[utf8x]{inputenc}
\usepackage{amsmath,amssymb}
\usepackage[hmargin=4cm,vmargin=4cm]{geometry}
\usepackage{algorithm2e}

\usepackage{xspace}
\usepackage{graphicx}

\newcommand{\field}[1]{\mathbb{#1}}
\newcommand{\GFZ}{\ensuremath{\field{F}_2}\xspace}
\newcommand{\ord}[1]{\ensuremath{\mathcal{O}\!\left(#1\right)}}
\newcommand{\kbar}{\ensuremath{\overline{k}}\xspace}

\newcommand{\Magma}{\textsc{Magma}\xspace}

\newcommand{\xBG}{\textsf{x86\_64}\xspace}
\newcommand{\Opteron}{\textsf{Opteron}\xspace}

\newcommand{\CT}{\textsf{Core 2}\xspace}
\newcommand{\Xeon}{\textsf{Xeon}\xspace}
\newcommand{\Sage}{Sage\xspace}
\newcommand{\PolyBoRi}{\textsc{PolyBoRi}\xspace}

\newcommand{\addrowsfromtable}{\textsc{AddRowsFromTable}}
\newcommand{\maketable}{\textsc{MakeTable}}
\newcommand{\gausssubmatrix}{\textsc{GaussSubmatrix}}
\newcommand{\plssubmatrix}{\textsc{PlsSubmatrix}}
\newcommand{\submatrix}[1]{\textsc{SubMatrix}(#1)}

\newcommand{\blind}[1]{#1}

\newenvironment{citeproof}{\vspace{0.3cm}\emph{Proof.}}{\vspace{0.3cm}}

\title{Efficient Decomposition of Dense Matrices over GF(2)}
\blind{
\author{Martin R.\ Albrecht\thanks{This author was supported by the Royal Holloway Valerie Myerscough Scholarship.}\inst{1} \and Cl{\'e}ment Pernet \inst{2}}
\institute{Information Security Group, Royal Holloway, University of London\\
Egham, Surrey TW20 0EX, United Kingdom\\ \email{M.R.Albrecht@rhul.ac.uk} \and
INRIA-MOAIS LIG, Grenoble Univ. ENSIMAG, Antenne de Montbonnot 51,\\
avenue Jean Kuntzmann, F-38330 MONTBONNOT SAINT MARTIN, France\\ \email{clement.pernet@imag.fr}}
}

\begin{document}

\maketitle

\begin{abstract}
In this work we describe an efficient implementation of a hierarchy of algorithms for the decomposition of dense matrices over the field with two elements ($\GFZ$). Matrix decomposition is an essential building block for solving dense systems of linear and non-linear equations and thus much research has been devoted to improve the asymptotic complexity of such algorithms. In this work we discuss an implementation of both well-known and improved algorithms in the M4RI library. The focus of our discussion is on a new variant of the M4RI algorithm -- denoted MMPF in this work -- which allows for considerable performance gains in practice when compared to the previously fastest implementation. We provide performance figures on \xBG CPUs to demonstrate the viability of our approach.
\end{abstract}

\section{Introduction}
We describe an efficient implementation of a hierarchy of algorithms for PLS decomposition of dense matrices over the field with two elements ($\GFZ$). The PLS decomposition is closely related to the well-known PLUQ and LQUP decompositions. However, it offers some advantages in the particular case of \GFZ. Matrix decomposition is an essential building block for solving dense systems of linear and non-linear equations (cf.~\cite{f4,courtois-klimov-patarin-shamir:eurocrypt2000}) and thus much research has been devoted to improve the asymptotic complexity of such algorithms. In particular, it has been shown that various matrix decompositions such as PLUQ, LQUP and LPS are essentially equivalent and can be reduced to matrix-matrix multiplication (cf.~\cite{jeannerod-pernet-storjohann:pluq2010}). Thus, we know that these decompositions can be achieved in $\ord{n^\omega}$ where $\omega$ is the exponent of linear algebra\footnote{For practical purposes we set $\omega = 2.807$.}. In this work we focus on matrix decomposition in the special case of $\GFZ$ and discuss an implementation of both well-known and improved algorithms in the M4RI library \cite{M4RI}. The M4RI library implements dense linear algebra over $\GFZ$ and is used by the \Sage \cite{sage} mathematics software and the \PolyBoRi~\cite{polybori} package for computing Gröbner bases. It is also the linear algebra library used in \cite{mxl2,mohamed-cabarcas-ding-buchmann-bulygin:icisc09}.

Our implementation focuses on 64-bit x86 architectures (\xBG), specifically the Intel \CT and the AMD \Opteron. Thus, we assume in this chapter that each native CPU word has 64 bits. However it should be noted that our code also runs on 32-bit CPUs and on non-x86 CPUs such as the PowerPC.

Element-wise operations over $\GFZ$ are relatively cheap compared to loads from and writes to memory. In fact, in this work we demonstrate that the two fastest implementations for dense matrix decomposition over $\GFZ$ (the one presented in this work and the one found in \Magma~\cite{magma} due to Allan Steel) perform worse for sparse matrices despite the fact that fewer field operations are performed. This indicates that counting raw field operations is not an adequate model for estimating the running time in the case of \GFZ.


This work is organised as follows. We will start by giving the definitions of reduced row echelon forms (RREF), PLUQ and PLS decomposition in Section~\ref{sec:pluq} and establish their relations. We will then discuss Gaussian elimination and the M4RI algorithm in Section~\ref{sec:gauss} followed by a discussion of cubic PLS decomposition and the MMPF algorithm in \ref{sec:mmpf}. We will then discuss asymptotically fast PLS decomposition in Section~\ref{sec:pluq-fast} and implementation issues in Section~\ref{sec:pluq-implementation}. We conclude by giving empirical evidence of the viability of our approach in Section~\ref{sec:pluq-results}.

\section{RREF and PLS}
\label{sec:pluq}

\begin{proposition}[PLUQ decomposition] 
Any $m \times n$ matrix A with rank $r$, can be written $A = P LU Q$ where $P$ and $Q$ are two permutation matrices, of dimension respectively $m \times m$ and $n \times n$, $L$ is $m \times r$ unit lower triangular and $U$ is $r \times n$ upper triangular.
\end{proposition}

\noindent
\begin{citeproof}
See \cite{jeannerod-pernet-storjohann:pluq2010}.
\end{citeproof}

\begin{proposition}[PLS decomposition]
\label{proposition:pls}
Any $m \times n$ matrix A with rank $r$, can be written $A = P L S$ where $P$ is a permutation matrix of dimension $m \times m$, $L$ is $m \times r$ unit lower triangular and $S$ is an $r \times n$ matrix which is upper triangular except that its columns are permuted, that is $S = UQ$ for $U$ $r \times n$ upper triangular and $Q$ is a $n \times n$ permutation matrix.
\end{proposition}

\begin{proof}
Write $A = PLUQ$ and set $S = UQ$. 
\end{proof}

Another way of looking at PLS decomposition is to consider the $A = LQUP$ decomposition \cite{IbMoHu82}. We have $A = LQUP = LSP$ where $S = QU$. We can also write $A = LQUP = SUP$ where $S=LQ$. Applied to $A^T$ we then get $A=P^TU^TS^T = P'L'S'$. Finally, a proof for Proposition~\ref{proposition:pls} can also be obtained by studying any one of the Algorithms \ref{alg:gausspluq}, \ref{alg:mmpf} or \ref{alg:pls}.

\begin{definition}[Row Echelon Form] An $m \times n$ matrix $A$ is in echelon form if all zero rows are grouped together at the last row positions of the matrix, and if the leading coefficient of each non zero row is one and is located to the right of the leading coefficient of the above row.
\end{definition}

\begin{proposition}
\label{proposition:rref}
Any $m \times n$ matrix can be transformed into echelon form by matrix multiplication.
\end{proposition}

\noindent
\begin{citeproof}
See \cite{jeannerod-pernet-storjohann:pluq2010}
\end{citeproof}

Note that while there are many PLUQ decompositions of any matrix $A$ there is always also a decomposition for which we have that $S = UQ^T$ is a row echelon form of $A$. In this work we compute $A = PLS$ such that $S$ is in row echelon form. Thus, a proof for Proposition~\ref{proposition:rref} can also be obtained by studying any one of the Algorithms \ref{alg:gausspluq}, \ref{alg:mmpf} or \ref{alg:pls}.

\begin{definition}[Reduced Row Echelon Form]
An $m \times n$ matrix $A$ is in reduced echelon form if it is in echelon form and each leading coefficient of a non zero row is the only non zero element in its column.
\end{definition}

\section{Gaussian Elimination and M4RI}
\label{sec:gauss}

Gaussian elimination 
is the classical, cubic algorithm for transforming a matrix into (reduced) row echelon form using elementary row operations only.
The ``Method of the Four Russians'' Inversion (M4RI) \cite{Ba08} reduces the number of additions required by Gaussian elimination by a factor of $\log n$ by using a caching technique inspired by Kronrod's method for matrix-matrix multiplication.

\subsection{The ``Method of the Four Russians'' Inversion (M4RI)}
The ``Method of the Four Russians'' inversion was introduced in \cite{Ba06} and later described in \cite{Ba07} and \cite{Ba08}. It inherits its name and main idea from the misnamed ``Method of the Four Russians'' multiplication \cite{ArDiKrFa70,AhHoUl74}.

To give the main idea consider for example the matrix $A$ of dimension $m \times n$ in Figure~\ref{fig:m4ri-visualisation}. The $k \times n$ ($k=3$) submatrix on the top has full rank and we performed Gaussian elimination on it. Now, we need to clear the first $k$ columns of $A$ for the rows below $k$ (and above the submatrix in general if we want the reduced row echelon form). There are $2^k$ possible linear combinations of the first $k$ rows, which we store in a table $T$. We index $T$ by the first $k$ bits (e.g., $0 1 1 \rightarrow 3$). Now to clear $k$ columns of row $i$ we use the first $k$ bits of that row as an index in $T$ and add the matching row of $T$ to row $i$, causing a cancellation of $k$ entries. Instead of up to $k$ additions this only costs one addition due to the pre-computation. Using Gray codes (or similar techniques) this pre-computation can be performed in $2^k$ vector additions and the overall cost is $2^k + m - k + k^2$ vector additions in the worst case (where $k^2$ accounts for the Gauss elimination of the $k \times n$ submatrix). The naive approach would cost $k \cdot m$ row additions in the worst case to clear $k$ columns. If we set $k = \log m$ then the complexity of clearing $k$ columns is $\ord{m + \log^2 m}$ vector additions in contrast to $\ord{m \cdot \log m}$ vector additions using the naive approach.

\begin{figure}[htbp]
\begin{align*}
A = \left(\begin{array}{rrr|rrrrrr}
1 & 0 & 0 & 1 & 0 & 1 & 1 & 1 & \dots \\
0 & 1 & 0 & 1 & 1 & 1 & 1 & 0 & \dots \\
0 & 0 & 1 & 0 & 0 & 1 & 1 & 1 & \dots \\
\dots \\
0 & 0 & 0 & 1 & 1 & 0 & 1 & 0 & \dots \\
\bf{1} & \bf{1} & \bf{0} & 0 & 1 & 0 & 1 & 1 & \dots \\
0 & 1 & 0 & 0 & 1 & 0 & 0 & 1 & \dots \\
\dots \\
\bf{1} & \bf{1} & \bf{0} & 1 & 1 & 1 & 0 & 1 & \dots 
\end{array}\right)
T = \left[\begin{array}{rrr|rrrrrr}
0 & 0 & 0 & 0 & 0 & 0 & 0 & 0 & \dots\\
0 & 0 & 1 & 0 & 0 & 1 & 1 & 1 & \dots \\
0 & 1 & 0 & 1 & 1 & 1 & 1 & 0 & \dots \\
0 & 1 & 1 & 1 & 1 & 0 & 0 & 1 & \dots \\
1 & 0 & 0 & 1 & 0 & 1 & 1 & 1 & \dots \\
1 & 0 & 1 & 1 & 0 & 0 & 0 & 0 & \dots \\
\bf{1} & \bf{1} & \bf{0} & 0 & 1 & 0 & 0 & 1 & \dots \\
1 & 1 & 1 & 0 & 1 & 1 & 1 & 0 & \dots \\
\end{array}\right]
\end{align*}
\label{fig:m4ri-visualisation}
\caption{M4RI Idea}
\end{figure}

This idea leads to Algorithm~\ref{alg:m4ri}. In this algorithm the subroutine \gausssubmatrix\ (cf.\ Algorithm~\ref{alg:gaussubmatrix}) performs Gauss elimination on a $k \times n$ submatrix of $A$ starting at position $(r,c)$ and searches for pivot rows up to $m$. If it cannot find a submatrix of rank $k$ it will terminate and return the rank $\kbar$ found so far. Note the technicality that the routine \gausssubmatrix\ and its interaction with  Algorithm~\ref{alg:m4ri} make use of the fact that all the entries in a column below a pivot are zero if they were considered already.

The subroutine \maketable\ (cf.\ Algorithm~\ref{alg:maketable}) constructs the table $T$ of all $2^k$ linear combinations of the $k$ rows starting a row $r$ and a column $c$, i.e.~it enumerates all elements of the vector space $\mathsf{span}(r,...,r+\kbar+1)$ spanned by the rows $r,\dots, r+\kbar-1$. Finally, the subroutine \addrowsfromtable\ (cf.\ Algorithm~\ref{alg:addrows}) adds the appropriate row from $T$ -- indexed by $k$ bits starting at column $c$ -- to each row of $A$ with index $i \not\in \{r,\dots,r+\kbar-1\}$. That is, it adds the  appropriate linear combination of the rows $\{r,\dots,r+\kbar-1\}$  onto a row $i$ in order to clear $k$ columns.

Note that the relation between the index $id$ and the row $j$ in $T$ is static and known \emph{a priori} because \gausssubmatrix\ puts the submatrix in reduced row echelon form. In particular this means that the $\kbar \times \kbar$ submatrix starting at $(r,c)$ is the identity matrix.

\begin{algorithm}
\KwIn{$A$ -- a $m \times n$ matrix}
\KwIn{$k$ -- an integer $k > 0$}
\KwResult{$A$ is in reduced row echelon form.}
\Begin{
$r,c \longleftarrow 0,0$\;
\While{$c<n$}{
   \lIf{$c+k > n$}{
     $k \leftarrow n - c$\;
   }
   $\kbar \longleftarrow$ \textsc{GaussSubmatrix}($A, r, c, k, m$)\;
   \If{$\kbar > 0$}{
     $T,L \longleftarrow$ \textsc{MakeTable}($A, r, c, \kbar$)\;
     \addrowsfromtable($A, 0, r, c, \kbar, T, L$)\;
     \addrowsfromtable($A, r+\kbar, m, c, \kbar, T, L$)\;
   }
   $r,c \longleftarrow r + \kbar, c + \kbar$\;
   \lIf{$k \neq \kbar$}{
     $c \leftarrow c + 1$\;
   }
}
}
\caption{M4RI}
\label{alg:m4ri}
\end{algorithm}

When studying the performance of Algorithm~\ref{alg:m4ri}, we expect the function \maketable\ to contribute most. Instead of performing $\kbar/2 \cdot 2^{\kbar} - 1$ additions \maketable\ only performs $2^{\kbar}-1$ vector additions. However, in practice the fact that $\kbar$ columns are processed in each loop iteration of \addrowsfromtable\ contributes signficiantly due to the better cache locality. Assume the input matrix $A$ does not fit into L2 cache. Gaussian elimination would load a row from memory, clear one column and likely evict that row from cache in order to make room for the next few rows before considering it again for the next column. In the M4RI algorithm more columns are cleared per load.

We note that our presentation of M4RI differs somewhat from that in \cite{Ba07}. The key difference is that our variant does not throw an error if it cannot find a pivot within the first $3k$ rows in \gausssubmatrix. Instead, our variant searches all rows and consequently the worst-case complexity is cubic. However, on average for random matrices we expect to find a pivot within $3k$ rows and thus expect the average-case complexity to be $\ord{n^3/\log n}$.

\section{M4RI and PLS Decomposition}
\label{sec:mmpf}

In order to recover the PLS decomposition of some matrix $A$, we can adapt Gaussian elimination to preserve the transformation matrix in the lower triangular part of the input matrix $A$ and to record all permutations performed.
This leads to Algorithm~\ref{alg:gausspluq} in the Appendix which modifies $A$ such that it contains $L$ in below the main diagonal, $S$ above the main diagonal and returns $P$ and $Q$ such that $P L S = A$ and $S Q^T = U$.

The main differences between Gaussian elimination and Algorithm~\ref{alg:gausspluq} are:
\begin{itemize}
 \item No elimination is performed above the currently considered row, i.e.~the rows $0,\dots,r-1$ are left unchanged. Instead elimination starts below the pivot, from row $r + 1$.
 \item Column swaps are performed at the end of Algorithm~\ref{alg:gausspluq} but not in Gaussian elimination. This step compresses $L$ such that it is lower triangular.
 \item Row additions are performed starting at column $r + 1$ instead of $r$ to preserve the transformation matrix $L$. Over any other field we would have to rescale $A[r,r]$ for the transformation matrix $L$ but over \GFZ this is not necessary.
\end{itemize}

\subsection{The Method of Many People Factorisation (MMPF)}
In order to use the M4RI improvement over Gaussian elimination for PLS decomposition, we have to adapt the M4RI algorithm.

\paragraph*{Column Swaps}
Since column swaps only happen at the very end of the algorithm we can modify the M4RI algorithm in the obvious way to introduce them.

\paragraph*{$U$ vs. $I$}
Recall, that the function \gausssubmatrix\ generates small $\kbar \times \kbar$ identity matrices. Thus, even if we  remove the call to the function \addrowsfromtable($A,0,r,c,\kbar,T$) from Algorithm~\ref{alg:m4ri} we would still eliminate up to $\kbar-1$ rows above a given pivot and thus would fail to produce $U$. The reason the original specification \cite{Ba06} of the M4RI requires  $\kbar \times \kbar$ identity matrices is to have a \emph{a priori} know\-ledge of the relationship between $id$ and $j$ in the function \addrowsfromtable. On the other hand the rows of any $\kbar \times n$ upper triangular matrix also form a basis for the $\kbar$-dimensional vector space $\textsf{span}(r,\dots,r+\kbar-1)$. Thus, we can adapt \gausssubmatrix\ to compute the upper triangular matrix instead of the identity. Then, in \textsc{MakeTable1} we can encode the actual relationship between a row $j$ of $T$ and $id$ in the lookup table $L$.

\paragraph*{Preserving $L$}
In Algorithm~\ref{alg:gausspluq} preserving the transformation matrix $L$ is straight forward: addition starts in column $c + 1$ instead of $c$. On the other hand, for M4RI we need to fix the table $T$ to update the transformation matrix correctly; For example, assume $\kbar = 3 $ and that the first row of the $\kbar \times n$ submatrix generated by \gausssubmatrix\ has the first $\kbar$ bits equal to \texttt{[1 0 1]}. Assume further that we want to clear $\kbar$ bits of a a row which also starts with \texttt{[1 0 1]}. Then -- in order to generate $L$ -- we need to encode that this row is cleared by adding the first row only, i.e. we want the first $\kbar = 3$ bits to be \texttt{[1 0 0]}. Recall that in the M4RI algorithm the $id$ for the row $j$ starting with \texttt{[1 0 0]} is \texttt{[1 0 0]} if expressed as a sequence of bits. Thus, to correct the table, we add the $\kbar$ bits of the \emph{a priori} $id$ onto the first $\kbar$ entries in $T$ (starting at $c$) as in \maketable\texttt{1}.

\paragraph*{Other Bookkeeping}
Recall that \gausssubmatrix's interaction with Algorithm~\ref{alg:m4ri} uses the fact that processed columns of a row are zeroed out to encode whether a row is ``done'' or not. This is not true anymore if we compute the PLS decomposition instead of the upper triangular matrix in \gausssubmatrix\ since we store $L$ below the main diagonal. Thus, we explicitly encode up to which row a given column is ``done'' in \plssubmatrix\ (cf.\ Algorithm~\ref{alg:pluqsubmatrix}). Finally, we have to take care not to include the transformation matrix $L$ when constructing $T$.

\begin{algorithm}[htbp]
\caption{\textsc{MakeTable1}}
\label{alg:maketable1}
\KwIn{$A$ -- a $m \times n$ matrix}
\KwIn{$r_{\textnormal{start}}$ -- an integer  $0 \leq r_{\textnormal{start}} < m$}
\KwIn{$c_{\textnormal{start}}$ -- an integer  $0 \leq c_{\textnormal{start}} < n$}
\KwIn{$k$ -- an integer $k > 0$}
\KwResult{Retuns an $2^k \times n$ matrix $T$ and the translation table $L$}
\Begin{
  $T \longleftarrow$ the $2^k \times n$ zero matrix\;
  \For{$1 \leq i < 2^k$}{
    $j \longleftarrow$ the row index of $A$ to add according to the Gray code\;
   add row $j$ of $A$ to the row $i$ of $T$ starting at $c_\textnormal{start}$\;
  }
  $L \longleftarrow$ an integer array with $2^k$ entries\;
  \For{$1 \leq i < 2^k$}{
    $id = \sum_{j=0}^{k} T[i,c_\textnormal{start} + j] \cdot 2^{k-j-1}$\;
    $L[id] \longleftarrow i$\;
  }

  \For{$1 \leq i < 2^k$}{
    $b_0,\dots,b_{\kbar-1} \longleftarrow$ bits of \emph{a priori} $id$ of the row $i$\;
   \For{$0 \leq j < \kbar$}{
     $T[i,c_{\textnormal{start}} + j] \longleftarrow T[i,c_{\textnormal{start}} + j] + b_j$\;
   }
  }
  \Return{$T,L$}\;
}
\end{algorithm}

These modifications lead to Algorithm~\ref{alg:mmpf} which computes the $PLS$ decomposition of $A$ in-place, that is $L$ is stored below the main diagonal and $S$ is stored above the main diagonal of the input matrix. Since none of the changes to the M4RI algorithm affect the asymptotical complexity, Algorithm~\ref{alg:mmpf} is cubic in the worst case and has complexity $\ord{n^3/\log n}$ in the average case.

\begin{algorithm}[h]
\KwIn{$A$ -- a $m \times n$ matrix}
\KwIn{$P$ -- a permutation vector of length $m$}
\KwIn{$Q$ -- a permutation vector of length $n$}
\KwIn{$k$ -- an integer $k > 0$}
\KwResult{PLS decomposition of $A$}
\SetKw{KwAnd}{and}
\Begin{
  $r,c \longleftarrow 0,0$\;
  \lFor{$0 \leq i<n$}{ $Q[i] \longleftarrow i$\; }
  \lFor{$0 \leq i<m$}{ $P[i] \longleftarrow i$\; }

  \While{$r < m$ \KwAnd $c < n$}{
    \lIf{$c + k > n$}{ $k \longleftarrow n - c$\; }
    $\kbar,d_r \longleftarrow $ \plssubmatrix($A, r, c, k, P, Q$)\;
  
    $U \longleftarrow$ the $\kbar \times n$ submatrix starting at $(r,0)$ where every entry prior to the upper triangular matrix starting at $(r,c)$ is zeroed out\;
  
    \eIf{$\kbar > 0$}{
      $T,L \longleftarrow$ \textsc{MakeTable1}($U, 0, c, \kbar$)\;
      \addrowsfromtable($A, d_r+1, m, c, \kbar, T, L$)\;
      $r,c \leftarrow r + \kbar, c + \kbar$\;
    }{
      \tcp{skip zero column}
      $c \leftarrow c + 1$\;
    }
  }
  \tcp{Now compress L}
  \lFor{$0 \leq j < r$}{
   swap the columns $j$ and $Q[j]$  starting at row $j$\;
  }
  \Return{$r$}\;
}
\caption{MMPF}
\label{alg:mmpf} 
\end{algorithm}


\section{Asymptotically Fast PLS Decomposition}
\label{sec:pluq-fast}
It is well-known that PLUQ decomposition can be accomplished in-place and in time complexity $\ord{n^\omega}$ by reducing it to matrix-matrix multiplication (cf.~\cite{jeannerod-pernet-storjohann:pluq2010}). We give a slight variation of the recursive algorithm from \cite{jeannerod-pernet-storjohann:pluq2010} in Algorithm~\ref{alg:pls}. We compute the PLS instead of the PLUQ decomposition.

\begin{algorithm}[htbp]
\KwIn{$A$ -- a $m \times n$ matrix}
\KwIn{$P$ -- a permutation vector of length $m$}
\KwIn{$Q$ -- a permutation vector of length $n$}
\KwResult{PLS decomposition of $A$}
\SetKw{KwAnd}{and}
\Begin{
  $n_0 \longleftarrow$ pick some integer $0 \leq n_0 < n$; \tcp{$n_0 \approx n/2$}
  $A_0 \longleftarrow$ \submatrix{$A,0,0,m,n_0$}\;
  $A_1 \longleftarrow$ \submatrix{$A,0,n_0,m,n$}\;
  $Q_0 \longleftarrow$ $Q[0,\dots,n_0]$\;

  $r_0 \longleftarrow$ \textsc{PLS}($A_0,P,Q_0$); \tcp{first recursive call}

  \lFor{$0 \leq i \leq n_0$}{$Q[i] \leftarrow Q_0[i]$\;}

  $A_{NW} \longleftarrow$ \submatrix{$A,   0,   0, r_0,   r_0$}\;
  $A_{SW} \longleftarrow$ \submatrix{$A, r_0,   0,   m,   r_0$}\;
  $A_{NE} \longleftarrow$ \submatrix{$A,   0, n_0, r_0,     n$}\;
  $A_{SE} \longleftarrow$ \submatrix{$A, r_0, n_0,   m,     n$}\;

  \If{$r_1$}{
      \tcp{Compute of the Schur complement}
      $A_1 \longleftarrow P \times A_1$\;
      $L_{NW} \longleftarrow$ the lower left triangular matrix in $A_{NW}$\; 
      $A_{NE} \longleftarrow L_{NW}^{-1} \times A_{NE}$\;
      $A_{SE} \longleftarrow A_{SE} + A_{SW} \times A_{NE}$\;
  }

   $P_1 \longleftarrow $ $P[r_0,\dots,m]$\;
   $Q_1 \longleftarrow $ $Q[n_0,\dots,n]$\;

   $r_1 \longleftarrow$ \textsc{PLS}($A_{SE},P_1,Q_1$); \tcp{second recursive call}

   $A_{SW} \longleftarrow P \times A_{SW}$\;
 
   \tcp{Update P \& Q}
   \lFor{$0 \leq i < m - r_0$}{$P[r_0 + 1] = P_1[i] + r_0$\;}
   \lFor{$0 \leq i < n - n_0$}{$Q[n_0 + i] \leftarrow Q_1[i] + n_0$\;}

   $j \leftarrow r_0$\;
   \lFor{$n_0 \leq i < n_0 + r_1$}{$Q[j] \leftarrow Q[i]$; $j \leftarrow j + 1$\;}

\tcp{Now compress L}
  $j \leftarrow n_0$\;
  \lFor{$r_0 \leq i < r_0+r_1$}{
   swap the columns $i$ and $j$  starting at row $i$\;
  }
  \Return{$r_0 + r_1$}\;
}
\caption{PLS Decomposition}
\label{alg:pls} 
\end{algorithm}

In Algorithm~\ref{alg:pls} the routine \submatrix{$r_s,c_s,r_e,c_e$} returns a ``view'' (cf.~\cite{matmulgf2}) into the matrix $A$ starting at row and column $r_s$ and $c_s$ resp. and ending at row and column $r_e$ and $c_e$ resp. We note that that the step $A_{NE} \longleftarrow L_{NW}^{-1} \times A_{NE}$ can be reduced to matrix-matrix multiplication (cf.~\cite{jeannerod-pernet-storjohann:pluq2010}). Thus Algorithm~\ref{alg:pls} can be reduced to matrix-matrix multiplication and has complexity $\ord{n^\omega}$. Since no temporary matrices are needed to perform the algorithm, except maybe in the matrix-matrix multiplication step, the algorithm is in-place.

\section{Implementation}
\label{sec:pluq-implementation}
Similarly to matrix multiplication (cf.\ \cite{matmulgf2}) it is beneficial to call Algorithm~\ref{alg:pls} until some ``cutoff'' bound and to switch to a base-case implementation (in our case Algorithm~\ref{alg:mmpf}) once this bound is reached. We perform the switch over if the matrix fits into 4MB or in L2 cache, whichever is smaller. These values seem to provide the best performance on our target platforms.

The reason we are considering the PLS decomposition instead of either the LQUP or the PLUQ decomposition is that the PLS decomposition has several advantages over $\GFZ$, in particular when the flat row-major representation is used to store entries.
\begin{itemize}
 \item We may choose where to cut with respect to columns in Algorithm~\ref{alg:pls}. In particular, we may choose to cut along word boundaries. For LQUP decomposition, where roughly all steps are transposed, column cuts are determined by the rank $r_0$.
 \item In Algorithm~\ref{alg:mmpf} rows are added instead of columns. Row operations are much cheaper than column operations in row-major representation.
 \item Column swaps do not occur in the main loop of either Algorithm ~\ref{alg:pls} or \ref{alg:mmpf}, but only row swaps are performed. Column swaps are only performed at the end. Column swaps are much more expensive than row swaps (see below).
 \item Fewer column swaps are performed for PLS decomposition than for PLUQ decomposition since U is not compressed.
\end{itemize}

One of the major bottleneck are column swaps. In Algorithm~\ref{alg:swap} a simple algorithm for swapping two columns $a$ and $b$ is given with bit-level detail. In Algorithm~\ref{alg:swap} we assume that the bit position of $a$ is greater than the bit position of $b$ for simplicity of presentation. The advantage of the strategy in Algorithm~\ref{alg:swap} is that it uses no conditional jumps in the inner loop, However, it still requires 9 instructions per row. On the other hand, we can add two rows with $9 \cdot 128 = 1152$ entries in 9 instructions if the SSE2 instruction set is available. Thus, for matrices of size $1152 \times 1152$ it takes roughly the same number of instructions to add two matrices as it does to swap two columns. If we were to swap every column with some other column once during some algorithm it thus would be as expensive as a matrix multiplication for matrices of these dimensions.

\begin{algorithm}[htbp]
\KwIn{$A$ -- a $m \times n$ matrix}
\KwIn{$a$ -- an integer $0 \leq a < b < n$}
\KwIn{$b$ -- an integer $0 \leq a < b < n$}
\KwResult{Swaps the columns $a$ and $b$ in $A$}
\SetKw{KwAnd}{and}
\Begin{
$M \longleftarrow$  the memory where $A$ is stored\;
$a_w, b_w \longleftarrow$ the word index of $a$ and $b$ in $M$\;
$a_b, b_b \longleftarrow$ the bit index of $a$ and $b$ in $a_w$ and $b_w$\;
$\Delta \longleftarrow$ $a_b - b_b$\;
$a_{m} \longleftarrow$ the bit-mask where only the $a_b$th bit is set to 1\;
$b_{m} \longleftarrow$ the bit-mask where only the $b_b$th bit is set to 1\;
\For{$0 \leq i < m$}{
  $R \longleftarrow$ the memory where the row $i$ is stored\;
  $R[a_w] \longleftarrow R[a_w] \oplus ((R[b_w] \odot b_{m}) >> \Delta)$\;
  $R[b_w] \longleftarrow R[b_w] \oplus ((R[a_w] \odot a_{m}) << \Delta)$\;
  $R[a_w] \longleftarrow R[a_w] \oplus ((R[b_w] \odot b_{m}) >> \Delta)$\;
}
}
\caption{Column Swap}
\label{alg:swap} 
\end{algorithm}

Another bottleneck for relatively sparse matrices in dense row-major representation is the search for pivots. Searching for a non-zero element in a row can be relatively expensive due to the need to identify the bit position. However, the main performance penalty is due to the fact that searching for a non-zero entry in one column is in a row-major representation is very cache unfriendly.

\begin{figure}[hbtp]
 \centering
 \includegraphics[width=0.9\textwidth]{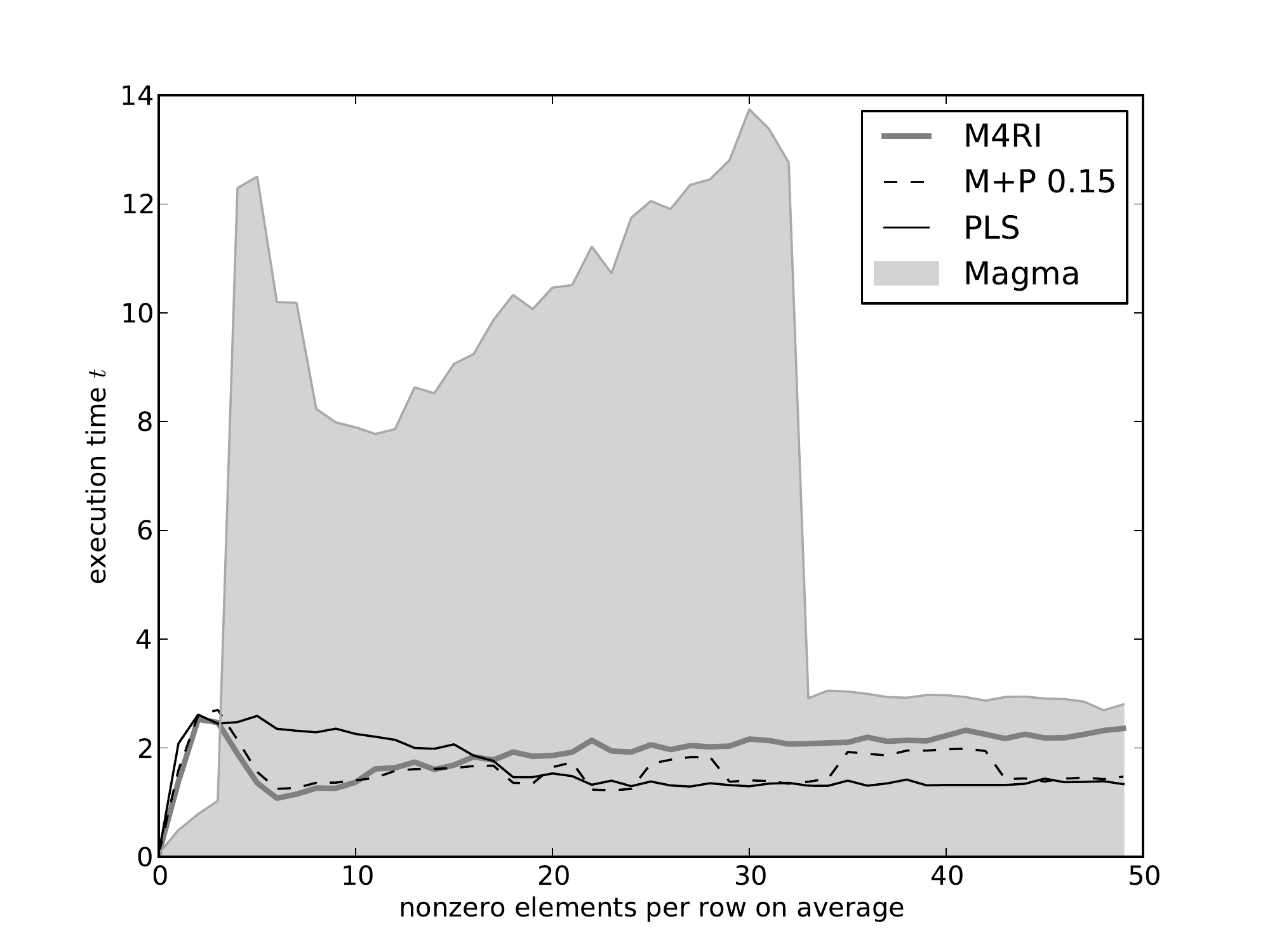}
 \caption{Sensitivity to density for $n=10^4$ on 2.6Ghz \Opteron}
 \label{fig:sparse-m4ri}
\end{figure}

Indeed, both our implementation and the implementation available in \Magma suffer from performance regression on relatively sparse matrices as shown in Figure~\ref{fig:sparse-m4ri}. We stress that this is despite the fact that the theoretical complexity of matrix decomposition is rank sensitive, that is, strictly less field operations have to be performed for low rank matrices. While the penalty for relatively sparse matrices is much smaller for our implementation than for \Magma, it clearly does not achieve the theoretical possible performance. Thus, we also consider 
a hybrid algorithm which starts with M4RI and switches to PLS-based elimination as soon as the (approximated) density reaches 15\%, denoted as `M+P 0.15'.

\section{Results}
\label{sec:pluq-results}
%

In Table~\ref{tab:pluq-random} we give average running time over ten trials for computing reduced row echelon forms of dense random $n \times n$ matrices over \GFZ. We compare the asymptotically fast implementation due to Allan Steel in \Magma, the cubic Gaussian elimination implemented by Victor Shoup in NTL, and both our implementations. Both the implementation in \Magma and our PLS decomposition reduce matrix decomposition to matrix multiplication. A discussion and comparison of matrix multiplication in the M4RI library and in \Magma can be found in \cite{matmulgf2}. In Table~\ref{tab:pluq-random} the column `PLS' denotes the complete running time for first computing the PLS decomposition and the computation of the reduced row echelon form from PLS. 


\begin{table}[htbp]
\begin{footnotesize}
\begin{center}
\begin{tabular}{|c|r|r|r|r||r|r|r|r|}
\hline
 & \multicolumn{4}{|c||}{64-bit Linux, 2.6Ghz \Opteron} & 
\multicolumn{4}{|c|}{64-bit Linux, 2.33Ghz \Xeon (E5345)}\\
\hline
$n$             &  {\Magma} &   {NTL} &   {M4RI} &    {PLS}& {\Magma} &   {NTL} &   {M4RI} &    {PLS}\\
             & {2.15-10} &   5.4.2 & 20090105 & 20100324& {2.16-7} &   5.4.2 & 20100324 & 20100324\\
\hline
$10,000$ &   3.351s &   18.45s &   2.430s &   1.452s&   2.660s &  12.05s &   1.360s &   0.864s\\
$16,384$ &  11.289s &   72.89s &  10.822s &   6.920s&   8.617s &  54.79s &   5.734s &   3.388s\\
$20,000$ &  16.734s &  130.46s &  19.978s &  10.809s&  12.527s & 100.01s &  10.610s &   5.661s\\
$32,000$ &  57.567s &  479.07s &  83.575s &  49.487s&  41.770s & 382.52s &  43.042s &  20.967s\\
$64,000$ & 373.906s & 2747.41s & 537.900s & 273.120s& 250.193s &      -- & 382.263s & 151.314s\\

\hline
\end{tabular}
\caption{RREF for random matrices}
\label{tab:pluq-random}
\end{center}
\end{footnotesize}
\end{table}

In Table~\ref{tab:pluq-practice} we give running times for matrices as they appear when solving non-linear systems of equations. The matrices HFE 25, 30 and 35 were contributed by Michael Brickenstein and appear during a Gröbner basis computation of HFE systems using \PolyBoRi. The Matrix MXL was contributed by Wael Said and appears during an execution of the MXL2 algorithm \cite{mxl2} for a random quadratic system of equations. We consider these matrices within the scope of this work since during matrix elimination the density quickly increases and because even the input matrices are dense enough such that we expect one non-zero element per 128-bit wide SSE2 XOR on average. The columns `M+P $0.xx$' denote the hybrid algorithms which start with M4RI and switch over to PLS based echelon form computation once the density of the remaining part of the matrix reaches 15\% or 20\% respectively.
We note that the relative performance of the M4RI and the PLS algorithm for these instances depends on particular machine configuration. To demonstrate this we give a set of timings for the Intel \Xeon X7460 machine \texttt{sage.math}\footnote{Purchased under National Science Foundation Grant No. DMS-0821725.} in Table~\ref{tab:pluq-practice}. Here, PLS always is faster than M4RI, while on a \Xeon E5345 M4RI wins for all HFE examples. We note that \Magma is not available on the machine \texttt{sage.math}.
The HFE examples show that the observed performance regression for sparse matrices does have an impact in practice and that the hybrid approach does look promising for these instances.

\begin{table}[htbp]
\begin{footnotesize}
\begin{center}
\begin{tabular}{|c|c|r|r|r|r|r|r|}
\hline
& & & \multicolumn{5}{|c|}{64-bit Fedora Linux, 2.33Ghz \Xeon (E5345)}\\
\hline
Problem & Matrix                 & Density &  Magma   &     M4RI &       PLS &  M+P 0.15 &  M+P 0.20\\
        & Dimension              &         &  2.16-7 &  20100324 &  20100324 &  20100429 &  20100429\\
\hline        
 HFE 25 & $12,307 \times 13,508$ &   0.076 &    3.68s &    1.94s &     2.09s &    2.33s &   2.24s\\
 HFE 30 & $19,907 \times 29,323$ &   0.067 &   23.39s &   11.46s &    13.34s &   12.60s &  13.00s\\
 HFE 35 & $29,969 \times 55,800$ &   0.059 &      --  &   49.19s &    68.85s &   66.66s &  54.42s\\
 MXL    & $26,075 \times 26,407$ &   0.185 &    55.15 &   12.25s &     9.22s &    9.22s &  10.22s\\
\hline
\hline
& & & \multicolumn{5}{|c|}{64-bit Ubuntu Linux, 2.66Ghz \Xeon (X7460)}\\
\hline
Problem & Matrix                 & Density & &      M4RI &       PLS &   M+P 0.15 & M+P 0.20\\
        & Dimension              &         & &  20100324 &  20100324 &   20100429 & 20100429\\
\hline        
 HFE 25 & $12,307 \times 13,508$ &   0.076 & &  2.24s &        2.00s &      2.39s &    2.35s\\
 HFE 30 & $19,907 \times 29,323$ &   0.067 & & 27.52s &       13.29s &     13.78s &    22.9s\\
 HFE 35 & $29,969 \times 55,800$ &   0.059 & &115.35s &       72.70s &     84.04s &  122.65s\\
 MXL    & $26,075 \times 26,407$ &   0.185 & & 26.61s &        8.73s &      8.75s &   13.23s\\
\hline
\hline
& & & \multicolumn{5}{|c|}{64-bit Debian/GNU Linux, 2.6Ghz \Opteron)}\\
\hline
Problem & Matrix                 & Density &  Magma   &     M4RI &        PLS &  M+P 0.15 &  M+P 0.20\\
        & Dimension              &         &  2.15-10 &  20100324 &  20100324 &  20100429 &  20100429\\
\hline        
 HFE 25 & $12,307 \times 13,508$ &   0.076 &   4.57s &    3.28s &       3.45s &   3.03s &   3.21s\\
 HFE 30 & $19,907 \times 29,323$ &   0.067 &  33.21s &   23.72s &      25.42s &  23.84s &  25.09s\\
 HFE 35 & $29,969 \times 55,800$ &   0.059 & 278.58s &  126.08s &     159.72s & 154.62s & 119.44s\\
 MXL    & $26,075 \times 26,407$ &   0.185 &  76.81s &   23.03s &      19.04s &  17.91s &  18.00s\\
\hline
\end{tabular}
\caption{RREF for matrices from practice.}
\label{tab:pluq-practice}
\end{center}
\end{footnotesize}
\end{table}

\section{Acknowledgments}
We would like to thank anonymous referees for helpful comments on how to improve our presentation.

\bibliographystyle{plain}
\bibliography{pluqm4ri_biblio}

\appendix
\clearpage
\section{Support Algorithms}

\begin{algorithm}
\KwIn{$A$ -- a $m \times n$ matrix}
\KwIn{$r_{\textnormal{start}}$ -- an integer 
$0 \leq r_{\textnormal{start}} < m$}
\KwIn{$r_{\textnormal{end}}$ -- an integer 
$0 \leq r_{\textnormal{start}} \leq r_{\textnormal{end}} < m$}
\KwIn{$c_{\textnormal{start}}$ -- an integer
 $0 \leq c_{\textnormal{start}} < n$}
\KwIn{$k$ -- an integer $k > 0$}
\KwIn{$T$ -- a $2^k \times n$ matrix}
\KwIn{$L$ -- an integer array of length $2^k$}
\Begin{
\For{$r_{\textnormal{start}} \leq i < r_{\textnormal{end}}$}{
  $id = \sum_{j=0}^{k} A[i,c_\textnormal{start} + j] \cdot 2^{k-j-1}$\;
  $j \longleftarrow L[id]$\;
  add row $j$ from $T$ to the row $i$ of $A$ starting at
column $c_\textnormal{start}$\;
} 
}
\caption{\textsc{AddRowsFromTable}}
\label{alg:addrows}
\end{algorithm}

\begin{algorithm}
\KwIn{$A$ -- a $m \times n$ matrix}
\KwIn{$r_{\textnormal{start}}$ -- an integer 
$0 \leq r_{\textnormal{start}} < m$}
\KwIn{$c_{\textnormal{start}}$ -- an integer
 $0 \leq c_{\textnormal{start}} < n$}
\KwIn{$k$ -- an integer $k > 0$}
\KwResult{Retuns an $2^k \times n$ matrix $T$}
\Begin{
$T \longleftarrow$ the $2^k \times n$ zero matrix\;
\For{$1 \leq i < 2^k$}{
  $j \longleftarrow$ the row index of $A$ to add according to the
Gray code\;
 add row $j$ of $A$ to the row $i$ of $T$ starting at
column $c_\textnormal{start}$\;
}
 $L \longleftarrow$ integer array allowing to index $T$ by $k$ bits
starting at column $c_{\textnormal{start}}$\;
\Return{$T,L$}\;
}
\caption{\textsc{MakeTable}}
\label{alg:maketable}
\end{algorithm}

\begin{algorithm}
\KwIn{$A$ -- a $m \times n$ matrix}
\KwIn{$r$ -- an integer $0 \leq r < m$}
\KwIn{$c$ -- an integer $0 \leq c < n$}
\KwIn{$k$ -- an integer $k > 0$}
\KwIn{$r_\textnormal{end}$ -- an integer $0 \leq r \leq r_\textnormal{end} <
m$}
\KwResult{Returns the rank $\kbar \leq k$ and puts the $\kbar
\times (n-c)$ submatrix starting at $A[r,c]$ in reduced row echelon form.}
\Begin{
  $r_s \longleftarrow r$\;
  \For{$c \leq j < c + k$}{
    $found \longleftarrow False$\;
    \For{$r_s \leq i < r_\textnormal{end}$}{
       \For(\tcp*[h]{clear the first columns}){$0 \leq l < j-c$}{
          \lIf{$A[i,c+l] \neq 0$}{
            add row $r+l$ to row $i$ of $A$ starting at column $c+l$\;
          }
       }
       \If(\tcp*[h]{pivot?}){$A[i,j] \neq 0$}{
         Swap the rows $i$ and $r_s$ in $A$\;
         \For(\tcp*[h]{clear above}){$r \leq l < r_s$}{
           \lIf{$A[l,j]\neq 0$}{
             add row $r_s$ to row $l$ in $A$ starting at column $j$\;
           }
         }
         $r_s \longleftarrow r_s + 1$\;
         $found \longleftarrow True$\;
         break\;
       }
    }
    \If{$found = False$}{
      \Return{j - c}\;
    }
  }
  \Return{j - c}\;
}
\caption{\textsc{GaussSubmatrix}}
\label{alg:gaussubmatrix}
\end{algorithm}

\begin{algorithm}
\caption{Gaussian PLS Decomposition}
\KwIn{$A$ -- a $m \times n$ matrix}
\KwIn{$P$ -- a permutation vector of length $m$}
\KwIn{$Q$ -- a permutation vector of length $n$}
\KwResult{PLS decomposition of $A$. Returns the rank of $A$.}
\SetKw{KwAnd}{and}
\SetKw{KwBreak}{break}
\Begin{
  $r, c \leftarrow 0,0$\;

  \While{$r<m$ \KwAnd $c < n$} {
    $found \longleftarrow False$\;
    \For(\tcp*[h]{search for some pivot}){$c \leq j < n$}{
      \For{$r \leq i < m$}{
        \lIf{$A[i,j]$}{
          $found \leftarrow True$ \KwAnd
          \KwBreak;
        }
      }
      \lIf{found}{\KwBreak;}
    }
    \eIf{found}{
      $P[r],Q[r] \longleftarrow i,j$\;
      swap the rows $r$ and $i$ in $A$\;      
      \tcp{clear below but preserve transformation matrix}
      \If{$j+1 < n$}{
        \For{$r+1 \leq l < m$}{
          \If{$A[l,j]$}{
            add the row $r$ to the row $l$ starting at column $j+1$\;
          }
        }
      }
      $r, c \longleftarrow r+1,j+1$\;
    }{\KwBreak;}
  }
  \lFor{$r \leq i <m$}{
    $P[i] \longleftarrow i$
  }\;
  \lFor{$r \leq i <n$}{
    $Q[i] \longleftarrow i$
  }\;
  \tcp{Now compress L}
  \lFor{$0 \leq j < r$}{
   swap the columns $j$ and $Q[j]$ starting at row $j$\;
  }
  \Return{$r$};
}
\label{alg:gausspluq}
\end{algorithm}

\begin{algorithm}[htbp]
\KwIn{$A$ -- a $m \times n$ matrix}
\KwIn{$s_r$ -- an integer $0 \leq s_r < m$}
\KwIn{$s_c$ -- an integer $0 \leq s_c < n$}
\KwIn{$k$ -- an integer $k > 0$}
\KwIn{$P$ -- a permutation vector of length $m$}
\KwIn{$Q$ -- a permutation vector of length $n$}
\SetKw{KwAnd}{and}
\SetKw{KwBreak}{break}
\KwResult{Returns the rank $\kbar \leq k$ and $d_r$ -- the last row considered.\\
Also puts the $\kbar \times (n-c)$ submatrix starting at $(r,c)$ in PLS decomposition form.}
\Begin{
  $done \longleftarrow$ all zero integer array of length $k$\;
  \For{$0 \leq r < k$}{
    $found \longleftarrow False$\;
    \For(\tcp*[h]{search for some pivot}){$s_r + r \leq i < m$}{
      \For(\tcp*[h]{clear before}){$0 \leq l < r$}{
        \If{$done[l] < i$} {
          \If{$A[i, s_c + l]\neq 0$}{
            add row $s_r+l$ to row $i$ in $A$ starting at column $s_c+l+1$\;
          }
          $done[l] \longleftarrow i$\;
        }
      }
      \If{$A[i, s_c + r] \neq 0$}{
        $found \longleftarrow True$\;
        \KwBreak\;
      }
    }
    \lIf{$found=False$}{
      \KwBreak\;
    }
    $P[s_r + r], Q[s_r + r] \longleftarrow i, s_c + r$\;
    swap the rows $s_r + r$ and $i$ in $A$\;
    $done[r] \longleftarrow i$\;
  }
  $d_r \longleftarrow \max(\{done[i] \mid i \in \{0,\dots,\kbar -1\}\})$\;
  \For(\tcp*[h]{finish submatrix}){$0 \leq c_2 < \kbar$ \KwAnd $r + c_2 < n -1$} {
    \For{$done[c_2] < r_2 \leq d_r$}{
      \If{$A[r_2, r + c_2]\neq 0$}{
        add row $r + c_2$ to row $r_2$ in $A$ starting at column $r + c_2 + 1$\;
      }
     }
  }  
  \Return{$r,d_r$}\;
} 
\caption{\plssubmatrix}
\label{alg:pluqsubmatrix}
\end{algorithm}


\end{document}